\documentclass{article}
\usepackage{graphicx}
\usepackage{amsmath}
\usepackage{braket}
\usepackage[linesnumbered,ruled,vlined]{algorithm2e}
\usepackage{threeparttable} 
\usepackage{url}

\title{Variational free complement method with Gaussian complements}
\author{Cong Wang \thanks{PO Box 26 Okemos, MI, 48805 (USA); congwang.webmail@gmail.com}}
\date{}

\newcommand{\G}{\mathrm{G}}

\begin{document}

\maketitle

\begin{abstract}
The complement functions in the free complement (FC) method are constructed by decontracting the Gaussian expansions of the Slater functions formed by the initial wavefunction and the $g$ functions. The helium ground state is used to demonstrate the accuracy. 
\end{abstract}

\section{Introduction}

Quantum mechanics provides a wide range of predictions for the behaviors of microscopic and macroscopic systems \cite{dirac1929quantum}. Numerical exact solutions are valuable for reference values of computational methods \cite{shiozaki2009higher} and for testing the accuracy of the underlying quantum theories \cite{drake2009high,mitroy2013theory,pachucki2024comprehensive}. Besides explicitly correlated approaches such as Hylleraas-type \cite{schwartz2006experiment,puchalski2006ground,sims2014hylleraas}, R12/F12 \cite{hattig2012explicitly,kong2012explicitly,ten2012explicitly}, and explicit correlated Gaussian (ECG) methods \cite{mitroy2013theory,bubin2013born}, the FC methods may provide a compact functional form towards the exact solution \cite{nakatsuji2004scaled,nakatsuji2005general,nakatsuji2012discovery}.     

Similar to the explicit correlated methods except ECG \cite{hattig2012explicitly,kong2012explicitly,ten2012explicitly,rychlewski2013explicitly,gruneis2017perspective}, the integrals in the FC methods exhibit computational challenges \cite{nakatsuji2012discovery} in the evaluations. The sampling method as the local Schr\"{o}dinger equation (LSE) approach \cite{nakatsuji2007solving, nakatsuji2012discovery,nakashima2013efficient,nakatsuji2015free, nakatsuji2018solving,nakatsuji2022direct,nakashima2023solving,nakashima2023potential}  has been proposed. In the meanwhile, the accuracy of the LSE is expected no longer second order of the wavefunction error \cite{nakatsuji2007solving} as an integral-based variational method does \cite{helgaker2000molecular,nakatsuji2015free}.

It has been recently proposed using $1 - e^{-\gamma r}$ as the scaling function $g$ in the FC method \cite{nakatsuji2022accurate,nakatsuji2024accurate,nakatsuji2024exact}. The results showed higher accuracies than the linear forms of the scaling functions with similar number of complement functions \cite{nakatsuji2022accurate,nakatsuji2024accurate,nakatsuji2024exact}. These advances bring a possibility of using the Gaussian functions to expand the Slater functions \cite{o1966gaussian,hehre1969self,hehre1970self, pietro1980molecular,pietro1981molecular,pietro1983molecular,lopez1987large,fernandez1988accurate,tew2005new, werner2007general} generated by the initial wavefunction $\psi_0$ and $g$ functions. Furthermore, if we decontract the linear combinations of the Gaussian functions to form the complement functions, this leads to an approach similar to the ECG method \cite{mitroy2013theory,bubin2013born} and it inherits the structure \cite{nakatsuji2004scaled,nakatsuji2005general,nakatsuji2012discovery} of the FC method.

Several features in the present Gaussian expansion approach may be anticipated:
\begin{itemize}
\item [(i)] describing the solution of a one-electron potential by the Gaussian functions exhibits an exponential type convergence of the electronic energy \cite{klopper1986gaussian, kutzelnigg1994theory,mckemmish2012gaussian,kutzelnigg2013expansion, bachmayr2014error,shaw2020completeness}. This rapid convergence suggests decontraction the linear combination of the Gaussian functions could provide flexibilities over using the Gaussian functions to evaluate specific integrals \cite{fernandez1988accurate,kurokawa2023gaussian};

\item [(ii)] the optimizations of the exponents in the ECG approaches \cite{mitroy2013theory} are  avoided to a large extent, except the optimizations of the Slater exponents \cite{nakatsuji2022accurate,nakatsuji2024exact}.

\item [(iii)] the geminal functions which contain all inter-electron pairs before the anti-symmetrization  appear in the high-order FC expansions in an $N$-electron system \cite{nakashima2013efficient}. Thus, the $N!$ cost of the antisymmetrization can be avoided at least at the low-order expansions \cite{nakashima2013efficient,nakatsuji2015solving,nakashima2020free}.

\end{itemize}

The present manuscript is organized as following. In Subsection \ref{sub_functional_form}, we describe the Gaussian expansions with the FC method. In Subsection \ref{sub_selection}, we explain in detail about using selection of the overlap integrals, based on Refs. \cite{von2008trapped,von2009correlated,rakshit2012hyperspherical,mitroy2013theory,Kalaee2014,Mosegaard2018,moriya2023novel,coomar2022quantum}, to reduce number of the complement functions and ameliorate the linear dependence issue. In Subsection \ref{subsection_im}, we then introduce the implementations of the present work. In Section \ref{results}, the numerical results and  discussions are provided. In Section \ref{summary_section}, the summary and outlook are presented.

\section{Methodology}
\label{method}
\subsection{Functional form of the FC wavefunction and Gaussian expansions} \label{sub_functional_form}
The FC wavefunction has the functional form \cite{nakatsuji2004scaled,nakatsuji2005general,nakatsuji2012discovery}
\begin{align}
   \psi_n = \Pi_{m=0}^{n-1} \left[ 1 + C_m g (H - E_m ) \right] \psi_0 \label{gH}
\end{align}
where $n$ characterizes the expansion order. $\{C_m\}$ are the parametrization coefficients.  $E_m$ is an energy in the evaluation. $g$ is a scaling function \cite{nakatsuji2004scaled,nakatsuji2005general,nakatsuji2012discovery} and $H$ is the Hamiltonian of the system. $\psi_n$ and $\psi_0$ are the $n$-th order and initial wavefunctions, respectively.

The computations of the FC methods are based on a linear combination of the complement functions $\{ \phi_i \}$
\begin{align}
   \psi_n = \sum_{i=1}^{M_n} c_i \phi_i \label{fc_expansion}
\end{align}
where $M_n$ is the number of the complement functions. The complement functions $\{ \phi_i \}$ are obtained by collecting linear independent terms from $\Pi_{m=0}^{n-1} g (H - E_m ) \psi_0$ and $\psi_0$.  $\{ c_i \}$ are the expansion coefficients, that can be determined  by either a variational method (with integral evaluations) \cite{nakatsuji2004scaled} or LSE (sampling, integral-free) \cite{nakatsuji2007solving}. The terms with diverged matrix elements are excluded \cite{nakatsuji2005analytically,kurokawa2005free,nakashima2007solving,kurokawa2008solving,nakashima2010solving,nakashima2011relativistic}. The completeness of such expansions in Eqs. \eqref{gH} and \eqref{fc_expansion} is related to Refs. \cite{klahn1977convergence_1,klahn1977convergence} and the generalizations \cite{hill1998completeness,wang2012variational}.

It maybe worthwhile to notice, the  value of $E_m$ in Eq. \eqref{gH} \cite{nakatsuji2012discovery}  may not be  specified from the FC theory \cite{nakatsuji2004scaled,nakatsuji2005general}. It could happen that some value of $E_m$ leads to cancellations of the terms between $H \psi_0$ and $-E_0 \psi_0$. For example, for the helium atom ground state, choose $\psi_0 = e^{- 27/16 (r_1 + r_2)}$ (the spin functions are omitted for the helium ground state and the atomic units \cite{szabo1996modern}  are used in the present work), $g=r_1 + r_2 + r_{12}$, and the electronic Hamiltonian from $\{r_1, r_2, r_{12}\}$ coordinates \cite{nakashima2007solving}. From this wavefunction, $E_0 = -729/256$ \cite{nakatsuji2005general}. From Eq. \eqref{gH}, $-E_0 \psi_0 = 729/256 \psi_0$ will cancel $-729/256 \psi_0$ generated from $( -1/2 \partial^2 / \partial r_1^2 - 1/2 \partial^2 / \partial r_2^2 ) \psi_0$.  No new complement functions will be generated. 

This issue may be avoided by choosing a different value of $E_m$,  only use the potential operators in generation the complement functions (the p-alone method) \cite{nakatsuji2022accurate}, or collect the linear independent terms separately from $gH \psi$ and $-gE_n \psi$ to avoid cancellations between each other.  In the present work, we choose the complement functions from the functional forms generated via the p-alone method \cite{nakatsuji2022accurate}. For example, the complement functions have the form for the helium ground state
\begin{align}
   (1 + P_{12} ) g_1^{n_1} g_2^{n_2} g_{12}^{n_{12}} \psi_0  \label{slater}
\end{align}
where $P_{12}$ is a permutation via electron number 1 and 2 \cite{nakatsuji2022accurate,nakatsuji2024accurate}. $g_1$, $g_2$, and $g_{12}$ are the $g$ functions. $g_1$ and $g_2$ contain the electron-nucleus distances $1$ and $2$. $g_{12}$ includes the electron-electron distance $12$. In the present work, we adopt 
\begin{align} 
g_1 &=  1 - e^{-\gamma_1 r_{1}} \label{g1} \\
g_2 &=  1 - e^{-\gamma_2 r_2}  \label{g2}  \\
g_{12} &= 1 - e^{-\gamma_{12} r_{12}}  \label{g12} 
\end{align}
where $\gamma_1$, $\gamma_2$, and $\gamma_{12}$ are the parameters that can be fixed by the cusp conditions \cite{kato1957eigenfunctions,pack1966cusp,nakatsuji2024accurate}. For a given order $n$, the new generated complement functions satisfy $n_1 + n_2 + n_{12} = n$ and $\psi_0 = e^{-\zeta (r_1 + r_2)}$ in Eq. \eqref{slater}.

A Slater function, including orbital and geminal functions, can be fitted as a linear combination of Gaussian functions \cite{o1966gaussian,hehre1969self,hehre1970self, pietro1980molecular,pietro1981molecular,pietro1983molecular,lopez1987large,fernandez1988accurate,tew2005new, werner2007general}, 
\begin{align}
  e^{-\zeta r} \approx \sum_{k=1}^{n_\G} c_k e^{-\alpha^{(k)} r^2} \label{gaussian_expansion}
\end{align}
here $r$ applies to one- and two-electron distances, including $r_1$, $r_2$, and $r_{12}$. The completeness of such expansion has been discussed in Refs. \cite{bukowski1995second,jeziorski1997completeness, hill1998completeness}.

In the STO-$n$G constructions \cite{o1966gaussian,hehre1969self,hehre1970self, pietro1980molecular,pietro1981molecular,pietro1983molecular,lopez1987large,fernandez1988accurate,tew2005new, werner2007general}, the exponents of the Slater and Gaussian functions are connected by \cite{o1966gaussian}
\begin{align}
  \frac{ \zeta'}{ \zeta}  =  \sqrt{\frac{  \alpha' }{ \alpha}}  \label{sto}
\end{align}
Notice the weighting functions in fitting the Slater geminal functions were used in Refs. \cite{tew2005new, werner2007general}. In the present work, we do not use weighting function. In Subsections \ref{sub_selection}, \ref{subsection_im}, and Section \ref{results}, we load the exponents of the Gaussian expansions of the one-electron Slater orbitals \cite{o1966gaussian,hehre1969self,fernandez1988accurate} for the Gaussian expansions of the two-particle geminal functions. Since the equations for fitting the Slater geminal functions \cite{tew2005new, werner2007general} without the weighting functions will have the same functional forms as the one-electron orbitals \cite{o1966gaussian,hehre1969self}, the aforementioned usage of the  orbital-expansion exponents for geminal functions is applicable.

Thus, for each Slater orbital and geminal function via the $1- e^{-\gamma r}$ type $g$ function \cite{nakatsuji2022accurate,nakatsuji2024accurate,nakatsuji2024exact} and the initial wavefunction in Eq. \eqref{gH}, we use the decontracted exponents $\{ \alpha^{(k)} \}$ from Eq. \eqref{gaussian_expansion} and the scaling relation Eq. \eqref{sto} to form the complement functions $\{ \phi_i \}$ in Eq. \eqref{fc_expansion}. The construction here is general. $r$ and $\gamma$ in the scaling function, $1- e^{-\gamma r}$ \cite{nakatsuji2022accurate,nakatsuji2024accurate,nakatsuji2024exact}, are for all one- and two-electron variables of a general $N$-electron system. Since the left-hand side (LHS) and the right-hand side (RHS) of Eq. \eqref{gaussian_expansion} have the same indices of the one- or two-electron distances, the efficient antisymmetrization method \cite{nakashima2013efficient} would be applicable for the Gaussian complement functions from Eq. \eqref{gaussian_expansion}.

The pseudocode for generating the Gaussian complement functions is in Algorithm \ref{alg_gen_ecg}, using Eqs. \eqref{slater} - \eqref{g12} as the example. Since Eq. \eqref{gaussian_expansion} is generally applicable for one- and two-electron distances \cite{o1966gaussian,hehre1969self,hehre1970self, pietro1980molecular,pietro1981molecular,pietro1983molecular,lopez1987large,fernandez1988accurate,tew2005new, werner2007general}, Algorithm \ref{alg_gen_ecg} with general antisymmetrizations \cite{nakashima2013efficient} is expected to be performed for various electronic systems.  

Notice in the input of Algorithm \ref{alg_gen_ecg}, for a general order $n$, the \texttt{fcs} are in the sequence of  $ e^{-\zeta (r_1 + r_2)}, ( 1 - e^{-\gamma_{12} r_{12}} ) e^{-\zeta (r_1 + r_2)}$, $ (1 + P_{12}) ( 1 - e^{-\gamma_1 r_1} ) e^{-\zeta (r_1 + r_2)}, \cdots$ . The permutation $P_{12}$ is applied to $n_1 \neq n_2$ in Eq. \eqref{slater}. In the output of Algorithm \ref{alg_gen_ecg},  $[ \alpha_1^{(1)}, \alpha_2^{(1)}, \alpha_{12}^{(1)}]$ and $[ \alpha_1^{(1)}, \alpha_2^{(1)}, \alpha_{12}^{(2)}]$ correspond to $e^{-\alpha_1^{(1)} r_1^2 - \alpha_2^{(1)} r_2^2 - \alpha_{12}^{(1)} r_{12}^2} $ and $e^{-\alpha_1^{(1)} r_1^2 - \alpha_2^{(1)} r_2^2 - \alpha_{12}^{(2)} r_{12}^2}$, respectively. No symmetrization of the spatial wavefunction nor $P_{12}$ is performed for the output of Algorithm \ref{alg_gen_ecg}. The reason is, the output of Algorithm \ref{alg_gen_ecg} is a bookkeeping  of all non-duplicated Gaussian complement functions. Symmetrizations of the Gaussian complement functions for the helium ground state are performed in Algorithm \ref{alg_selection}.

The technical specifications and realizations of Algorithm \ref{alg_gen_ecg} can be achieved by the adopted software. For the removals of the duplicated entries in lines 4 and 9, we adopt the default threshold of the \texttt{math.isclose} function \cite{python_math} of \texttt{Python} \cite{van1991interactively}  version 3.12.3. Namely, the relative error is not larger than $1\times 10^{-9}$ \cite{python_math}.  We adopt the symbolic library  \texttt{SymPy} \cite{10.7717/peerj-cs.103} version 1.14.0 for the expansions \cite{sympy_expand} in line 3. This specifies the order of the terms from expanding the product of the $g$ functions, Eqs. \eqref{g1} - \eqref{g12}. For instance, by adopting the parameters $\zeta = 1.6875$ and $\gamma_{12} =0.5$ as Ref. \cite{nakatsuji2024accurate}, $(1.0- e^{- 0.5 r_{12}})^2 e^{-1.6875 r_1  - 1.6875 r_2}$ forms $1.0e^{- 1.6875 r_1 -1.6875 r_2} + 1.0e^{- 1.6875r_1 -1.0 r_{12} - 1.6875 r_2}  -2.0 e^{-1.6875r_1 -0.5 r_{12} -1.6875 r_2}$ and $[1.6875, 1.6875,0.]$, $[1.6875, 1.6875, 1.0], [1.6875, 1.6875, 0.5]$. 
Use the \texttt{product} function in \texttt{itertools} \cite{python_product} of  \texttt{Python} \cite{van1991interactively}  version 3.12.3 in line 7 for the Cartesian products will specify the element order in the list as varying the rightmost index first \cite{python_product}. 

In addition, in line 6 of Algorithm  \ref{alg_gen_ecg}, we expand the Slater functions formed by the initial wavefunction $\psi_0$ and the $g$ functions Eqs. \eqref{g1} - \eqref{g12}, e.g., $\zeta+\gamma_1$ in $[\zeta+\gamma_1, \zeta, 0. ]$ from $(1+ P_{12})(1 - e^{-\gamma_1 r_1}) e^{-\zeta (r_1 + r_2)}$ and $2\gamma_{12}$ from $(1 - e^{-\gamma_{12} r_{12}})^2 e^{-\zeta (r_1 + r_2)}$. For the non-explicit correlated Slater exponent, e.g., 0. in $[\zeta, \zeta, 0.]$,  the single entry of the Gaussian exponent 0. is returned from the expansion Eq. \eqref{gaussian_expansion}. No coefficients $\{ c_k \}$ in Eq. \eqref{gaussian_expansion} is used, since we use the decontracted Gaussian functions. The notation $\leftarrow  \cup$ \cite{EricksonAlgorithms} for appending an element into a list is from Ref. \cite{EricksonAlgorithms}.  $\Leftarrow$ in Algorithm 1 means assigning the values from the RHS to the format of the LHS.

\begin{algorithm}[h] 
    \caption{Generating Gaussian exponents from complement functions consist with Slater functions (orbitals and geminals)}  \label{alg_gen_ecg}
    \KwIn{Complement functions \texttt{fcs} generated at a given order $n$ of the FC method }
    \KwOut{A list of exponents for Gaussian complement functions as \texttt{unselected\_gfcs},   $[[\alpha_1^{(1)}, \alpha_2^{(1)}, \alpha_{12}^{(1)}], [\alpha_1^{(1)}, \alpha_2^{(1)}, \alpha_{12}^{(2)}], \cdots]$.  The superscripts $(1)$ and $(2)$ correspond to $k=1$ and $k=2$ in Eq. \eqref{gaussian_expansion}, respectively.}
    \texttt{exp\_coeffs\_fcs} = [ ]\;
    \For{ \textnormal{\texttt{fc}} in\textnormal{ \texttt{fcs}}}{
       $[\zeta_1, \zeta_2, \zeta_{12}] \Leftarrow   $ Eq. \eqref{slater}, e.g.,
        \begin{itemize}
            \item[] $[\zeta, \zeta, 0. ] \Leftarrow  e^{-\zeta (r_1 + r_2) }$,
            \item[] $[\zeta, \zeta, 0. ]$, $[\zeta, \zeta, \gamma_{12} ] \Leftarrow  (1 - e^{- \gamma_{12} r_{12}}) e^{ - \zeta (r_1 + r_2) } $, 
            \item[] $[\zeta+\gamma_1, \zeta, 0. ]$, $[\zeta, \zeta, 0. ]$  $\Leftarrow $ $ ( 1 + P_{12} )(1 - e^{-\gamma_1 r_1}) e^{-\zeta (r_1 + r_2) } $,  where $\zeta+\gamma_1$ and $\zeta$ are in descending order\;
        \end{itemize}
        \texttt{exp\_coeffs\_fcs}  $ \leftarrow  \texttt{exp\_coeffs\_fcs}  \cup [\zeta_1, \zeta_2, \zeta_{12}] $\; 
        } 
    Remove duplicated entries in \texttt{exp\_coeffs\_fcs}\;

    \For {\textnormal{\texttt{exp\_coeffs\_fc}} in \textnormal{\texttt{exp\_coeffs\_fcs} }  }{
        Expand each exponent in \texttt{exp\_coeffs\_fc} according to the STO-$n$G expansions \cite{o1966gaussian,hehre1969self,fernandez1988accurate} at $\zeta = 1$ in Eq. \eqref{gaussian_expansion} and the scaling relation Eq. \eqref{sto} from $\zeta = 1$ to each exponent in \texttt{exp\_coeffs\_fc}. The formed exponents of the STO-$n$G expansions are listed in descending order\;
        Form the Cartesian products of the Gaussian exponents of $r_1$, $r_2$, and $r_{12}$  as the Gaussian complement exponents\; 
        Loop over the Cartesian products of the Gaussian-complement exponents, sort the exponents of $r_1$ and $r_2$ in descending order\; 
        Remove the duplicated Gaussian complement functions\; 
        \texttt{unselected\_gfcs} $\leftarrow$ \texttt{unselected\_gfcs} $\cup$ the non-duplicated Gaussian complement functions\;
        }
    \Return  \texttt{unselected\_gfcs} 
\end{algorithm}

\subsection{Selection algorithm based on the overlap integral criteria} \label{sub_selection}
The decontracted Gaussian expansion described in Subsection \ref{sub_functional_form} can generate a large number of Gaussian complement functions and linear dependence issue in the variational calculations. We then adopt the selection method based on the overlap integrals in Refs. \cite{von2008trapped,von2009correlated,rakshit2012hyperspherical,mitroy2013theory,Kalaee2014,Mosegaard2018,moriya2023novel,coomar2022quantum}.  The  pseudocode, using the helium ground state as the example and with the specifications of the selection process, is in Algorithm \ref{alg_selection}. In Algorithm \ref{alg_selection}, a unique output of the selected Gaussian complement functions is generated from a unique input.

In addition, in lines 10, 12, and 13 of Algorithm \ref{alg_selection}, the symmetrizations are performed on numerical distinct exponents of $r_1$ and $r_2$. Namely, the absolute value of the difference of the exponents of $r_1$ and $r_2$ is larger than a given threshold. Here, $1\times 10^{-12}$ is used. 

Similar to Algorithm \ref{alg_gen_ecg}, using  antisymmetrizations for a general $N$-electron systems \cite{nakashima2013efficient} would allow Algorithm \ref{alg_selection} on various electronic systems.

\begin{algorithm}[h] 
    \caption{Selection of Gaussian complement functions based on overlap matrix values} \label{alg_selection}
    \KwIn{List of  exponents of Gaussians complement functions, \texttt{unselected\_gfcs}, from Algorithm \ref{alg_gen_ecg}  }
    \KwOut{A list of complement functions generated by Gaussian expansions}
    \texttt{duplicated\_gfcs} = []  \;
    $M_n^{\mathrm{before}} \leftarrow $  the  number of complement functions for the input ($\phi_{\G}^{\mathrm{before}}$)\;
    \For{$i = 1$ to $M_n^{\mathrm{before}}$}{
        \If{ $i \in$  \textnormal{\texttt{duplicated\_gfcs}} }{ skip\; }
        \For{$j = i+1$ to $M_n^{\mathrm{before}}$}{
        \If{ $j \in$  \textnormal{\texttt{duplicated\_gfcs}} }{ skip\;}
\If{  $\frac{ \braket{\tilde{\phi}^{\mathrm{before}}_{\G,i} |\tilde{\phi}^{\mathrm{before}}_{\G,j}}}{ \lVert   \tilde{\phi}^{\mathrm{before}}_{\G, i}  \rVert \lVert   \tilde{\phi}^{\mathrm{before}}_{\G, j}  \rVert  } > \mathrm{overlap\,\, threshold}$ \textup{\cite{von2008trapped,Kalaee2014,Mosegaard2018}}   }{  \texttt{duplicated\_gfcs} $\leftarrow$ \texttt{duplicated\_gfcs}  $\cup$  $j$.  Here, $\tilde{}$ means symmetrized function.  $ \lVert \,  \rVert$ represents the $L^2$ norm, i.e., $ \lVert \phi \rVert := \sqrt{\braket{\phi| \phi} }$ \cite{klahn1977convergence_1}\;  }   } 
        }
    Collect \{\texttt{unselected\_gfcs}  $ \not \in$ \texttt{duplicated\_gfcs}\}\;
     Symmetrize \texttt{unselected\_gfcs}\;  
     \Return Form complement functions from symmetrized \texttt{unselected\_gfcs}: $ (1 + P_{12} ) e^{-\alpha_1^{(1)} r_1^2- \alpha_2^{(1)}  r_2^2 - \alpha_{12}^{(1)}  r_{12}^2 }, \cdots $
\end{algorithm}

\subsection{Implementations} \label{subsection_im}
In the present work, we use the \texttt{mpmath} multi-precision library version 1.3.0 \cite{mpmath} in the \texttt{Python} program language \cite{van1991interactively} version 3.12.3 with 50 decimal precision. The evaluations of the ECG integrals are based on the formulae in Ref. \cite{mitroy2013theory}. In addition, we have implemented the FC method based on Eqs. \eqref{slater} - \eqref{g12} \cite{nakatsuji2022accurate,nakatsuji2024accurate} without the Gaussian expansion, Eq. \eqref{gaussian_expansion}, for the comparison in Section \ref{results}. The formulae in Ref. \cite{harris2004singular} for the integrals have been adopted. The \texttt{SymPy} symbolic library \cite{10.7717/peerj-cs.103} version 1.14.0 is used for the symbolic computations of the complement functions, e.g., the formation and expansions \cite{sympy_expand} of $(1- e^{-\gamma_{12} r_{12}})^2 e^{-\zeta (r_1 + r_2)}$ for Algorithms \ref{alg_gen_ecg} and \ref{alg_selection} with $\zeta = 1.6875$ as in Ref. \cite{nakatsuji2024accurate} and in obtaining the variables in the integral formulae in Refs. \cite{mitroy2013theory,harris2004singular}.

To enhance the efficiency, the \texttt{Julia} program language \cite{Julia-2017} version 1.11.6 is used as the generalized eigenvalue solver with the \texttt{BigFloat} type for 50 decimal precision. Specifically, we have implemented the canonical orthogonalization to solve a generalized eigenvalue problem \cite{szabo1996modern}. The functions associated to the eigenvalues of the overlap matrix from normalized bases smaller than $1 \times 10^{-30}$ will be dropped. The \texttt{eigen} in \texttt{GenericLinearAlgebra} version 0.3.18 and matrix multiplication of  \texttt{Julia} \cite{Julia-2017}  are used. Notice in solving the generalized eigenvalue problem, the matrix elements are evaluated via the normalized complement functions (symmetrized with numerically distinct exponents for $n_1 \neq n_2$ with input of Algorithm \ref{alg_gen_ecg}  described in Subsection \ref{sub_functional_form}  and $\alpha_1$ and $\alpha_2$  of
Algorithm \ref{alg_selection}  described in Subsection \ref{sub_selection})  have been normalized, as suggested in Ref. \cite{bubin2008energy}.

The \texttt{cursor} code editor associated with large language models \cite{cursor} has been used in generating codes.

\section{Results and discussions} \label{results}
We set the parameters $\{\gamma_1, \gamma_2, \gamma_{12} \}$ in Eqs. \eqref{g1} - \eqref{g12} according to the cusp conditions \cite{kato1957eigenfunctions,pack1966cusp}, as proposed in Ref. \cite{nakatsuji2024accurate}. For the helium ground state, $\zeta=1.6875$, $\gamma_1 = \gamma_2 = 0.3125$, and $\gamma_{12} = 0.5$ are adopted as in Ref. \cite{nakatsuji2024accurate}. The permutation for different exponents of electron 1 and 2 are applied, as described in Algorithm \ref{alg_gen_ecg}.  The results are presented in Tables \ref{tab1} and \ref{tab2}.

\begin{table}[h]
\caption{Results of Gaussian expanded variational calculations compared with the FC method of $g$ function, Eqs. \eqref{g1} - \eqref{g12}, \cite{nakatsuji2022accurate,nakatsuji2024accurate} without the Gaussian expansions Eq. \eqref{gaussian_expansion} for the ground state of the helium atom. $\psi_0 = e^{-\zeta (r_1 + r_2) }$, $\zeta = 1.6875$ \cite{nakatsuji2024accurate}. 0.95 and 0.98 are the overlap thresholds in Algorithm \ref{alg_selection}, that is used in screening the linear dependent Gaussian complement functions.  $n\G = 3$, $n\G = 6$, $n\G = 10$, and $n\G = 14$ correspond to the decontracted STO-3G  \cite{hehre1969self}, STO-6G \cite{hehre1969self}, STO-10G \cite{o1966gaussian}, and STO-14G \cite{fernandez1988accurate}, respectively. $M_n^{\mathrm{before}}$ and $M_n^{\mathrm{after}}$ are denoted to the number  of the complement functions before (already without duplicated functions via Algorithm \ref{alg_gen_ecg}) and after selection for the values of overlap matrices, respectively.  $s_{\mathrm{min}}$ stands for the minimum value of the overlap matrix. $E$ stands for the electronic energy \cite{szabo1996modern}. Energies and $s_{\mathrm{min}}$ are rounded to even \cite{python_format,python_round,goldberg1991every}. }

\begin{threeparttable}
\begin{tabular}{ccccccccc} \hline
\multicolumn{3}{c}{} & \multicolumn{3}{c}{0.95} & \multicolumn{3}{c}{0.98} \\
$n$ &  $n\G$ &  $M_n^{\mathrm{before}}$ &  $M_n^{\mathrm{after}}$  &  $s_{\mathrm{min}}$ & $ E$ & $M_n^{\mathrm{after}}$  & $s_{\mathrm{min}}$  &  $ E$   \\ \hline
0 & 3 & 6 & 6 & 2.1 $\times$ 10$^{-2}$ & -2.831550 & 6 & 2.1 $\times$ 10$^{-2}$ & -2.831550 \\
 & 6 & 21 & 21 & 1.5 $\times$ 10$^{-4}$ & -2.877298 & 21 & 1.5 $\times$ 10$^{-4}$ & -2.877298 \\
 & 10 & 55 & 55 & 1.2 $\times$ 10$^{-5}$ & -2.879007 & 55 & 1.2 $\times$ 10$^{-5}$ & -2.879007 \\
 & 14 & 105 & 92 & 5.6 $\times$ 10$^{-7}$ & -2.879018 & 103 & 1.8 $\times$ 10$^{-8}$ & -2.879026 \\
1 & 3 & 33 & 9 & 7.3 $\times$ 10$^{-3}$ & -2.852241 & 14 & 1.5 $\times$ 10$^{-3}$ & -2.859192 \\
 & 6 & 183 & 37 & 1.5 $\times$ 10$^{-4}$ & -2.901624 & 74 & 8.9 $\times$ 10$^{-6}$ & -2.902425 \\
 & 10 & 705 & 106 & 1.1 $\times$ 10$^{-5}$ & -2.903621 & 227 & 3.3 $\times$ 10$^{-7}$ & -2.903712 \\
 & 14 & 1771 & 170 & 5.5 $\times$ 10$^{-7}$ & -2.903475 & 361 & 6.0 $\times$ 10$^{-11}$ & -2.903722 \\
2 & 3 & 93 & 18 & 2.5 $\times$ 10$^{-3}$ & -2.876569 & 26 & 7.5 $\times$ 10$^{-5}$ & -2.876192 \\
 & 6 & 582 & 65 & 9.5 $\times$ 10$^{-5}$ & -2.902937 & 112 & 6.0 $\times$ 10$^{-7}$ & -2.902762 \\
 & 10 & 2410 & 186 & 5.3 $\times$ 10$^{-6}$ & -2.903656 & 335 & 1.7 $\times$ 10$^{-8}$ & -2.903718 \\
 & 14 & 6286 & 231 & 5.8 $\times$ 10$^{-8}$ & -2.903503 & 507 & 4.6 $\times$ 10$^{-11}$ & -2.903723 \\
3 & 3 & 201 & 23 & 6.7 $\times$ 10$^{-4}$ & -2.880483 & 49 & 6.4 $\times$ 10$^{-6}$ & -2.892051 \\
 & 6 & 1338 & 90 & 1.4 $\times$ 10$^{-5}$ & -2.903071 & 186 & 7.3 $\times$ 10$^{-8}$ & -2.903416 \\
 & 10 & 5710 & 235 & 6.1 $\times$ 10$^{-7}$ & -2.903670 & 487 & 7.3 $\times$ 10$^{-9}$ & -2.903722 \\
 & 14 & 15106 & 295 & 5.4 $\times$ 10$^{-8}$ & -2.903510 & 694 & 4.2 $\times$ 10$^{-11}$ & -2.903724\tnote{a} \\
0 & Slater\tnote{b}  & \multicolumn{2}{c}{1\tnote{b}}  & 1.0 $\times$ 10$^{0}$ & -2.847656\tnote{b}  &  &    \\
1 & Slater\tnote{c}  & \multicolumn{2}{c}{3\tnote{c}}  & 2.4 $\times$ 10$^{-2}$ & -2.893591\\ 
2 & Slater\tnote{c}  & \multicolumn{2}{c}{7\tnote{c}} & 7.5 $\times$ 10$^{-4}$ & -2.903095 \\
3 & Slater\tnote{c}  & \multicolumn{2}{c}{13\tnote{c}} & 2.7 $\times$ 10$^{-5}$ & -2.903629 \\
 &  ECG\tnote{d} & \multicolumn{2}{c}{100\tnote{d}}   &    &  -2.903723818\tnote{d} \\
\multicolumn{1}{l}{$E_{\mathrm{ref}}$}  & \multicolumn{8}{r}{-2.903724377\tnote{e}   } \\
\hline
\end{tabular} \label{tab1}
    \begin{tablenotes}
      \item[a] -2.903723829. 
      \item[b] Without the Gaussian expansion, Eq. \eqref{gaussian_expansion}. No removal of the complement functions based on the overlap integrals. The same applies to all combined columns under $M_n^{\mathrm{before}}$ and  $M_n^{\mathrm{after}}$. The electronic energy is of Ref. \cite{nakatsuji2004scaled}.
      \item[c] The number of the complement functions is the same as Ref. \cite{nakatsuji2024accurate}. In Ref. \cite{nakatsuji2024accurate}, the exponents of the Slater orbitals are further optimized. In the present work, $\zeta = 1.6875$ remains the same.  
      \item[d] ECG exponents optimized individually \cite{rybak1989accurate}.
      \item[e] Refs. \cite{schwartz2006further,nakashima2007solving,kurokawa2008solving,nakashima2008accurately}.
    \end{tablenotes}
\end{threeparttable}
\end{table}

\begin{table}[h]
\caption{Results of Gaussian expanded variational calculations for the ground state of the helium atom. $\psi_0 = e^{-\zeta (r_1 + r_2) }$, $\zeta = 1.6875$.  $n\G = 3$, $n\G = 6$, $n\G = 10$, and $n\G = 14$ correspond to the decontracted STO-3G  \cite{hehre1969self}, STO-6G \cite{hehre1969self}, STO-10G \cite{o1966gaussian}, and STO-14G \cite{fernandez1988accurate}, respectively. $M_n^{\mathrm{before}}$ and $M_n^{\mathrm{after}}$ are denoted to the number  of the complement functions before (already without duplicated functions via Algorithm \ref{alg_gen_ecg}) and after selection for the values of overlap matrices.  0.99 and 0.995 are the overlap thresholds in Algorithm \ref{alg_selection}, that is used in screening the linear dependent Gaussian complement functions. $s_{\mathrm{min}}$ stands for the minimum value of the overlap matrix. $E$ stands for the electronic energy \cite{szabo1996modern}. Energies and $s_{\mathrm{min}}$ are rounded to even \cite{python_format,python_round,goldberg1991every}.}
\begin{threeparttable}
\begin{tabular}{ccccccccc} \hline
\multicolumn{3}{c}{} & \multicolumn{3}{c}{0.99} & \multicolumn{3}{c}{0.995} \\
$n$ &  $n\G$ &  $M_n^{\mathrm{before}}$ &  $M_n^{\mathrm{after}}$  &  $s_{\mathrm{min}}$ & $ E$ & $M_n^{\mathrm{after}}$  & $s_{\mathrm{min}}$  &  $ E$   \\ \hline
0 & 3 & 6 & 6 & 2.1 $\times$ 10$^{-2}$ & -2.831550 & 6 & 2.1 $\times$ 10$^{-2}$ & -2.831550 \\
 & 6 & 21 & 21 & 1.5 $\times$ 10$^{-4}$ & -2.877298 & 21 & 1.5 $\times$ 10$^{-4}$ & -2.877298 \\
 & 10 & 55 & 55 & 1.2 $\times$ 10$^{-5}$ & -2.879007 & 55 & 1.2 $\times$ 10$^{-5}$ & -2.879007 \\
 & 14 & 105 & 105 & 5.6 $\times$ 10$^{-9}$ & -2.879026 & 105 & 5.6 $\times$ 10$^{-9}$ & -2.879026 \\
1 & 3 & 33 & 20 & 2.0 $\times$ 10$^{-4}$ & -2.870733 & 22 & 7.0 $\times$ 10$^{-5}$ & -2.871656 \\
 & 6 & 183 & 95 & 2.8 $\times$ 10$^{-7}$ & -2.902574 & 113 & 6.2 $\times$ 10$^{-8}$ & -2.902592 \\
 & 10 & 705 & 290 & 1.2 $\times$ 10$^{-8}$ & -2.903717 & 346 & 1.9 $\times$ 10$^{-9}$ & -2.903718 \\
 & 14 & 1771 & 514 & 2.5 $\times$ 10$^{-12}$ & -2.903723 & 679 & 1.2 $\times$ 10$^{-13}$ & -2.903723 \\
2 & 3 & 93 & 41 & 4.1 $\times$ 10$^{-6}$ & -2.889258 & 53 & 2.9 $\times$ 10$^{-7}$ & -2.891079 \\
 & 6 & 582 & 179 & 1.5 $\times$ 10$^{-9}$ & -2.903294 & 265 & 9.1 $\times$ 10$^{-11}$ & -2.903308 \\
 & 10 & 2410 & 522 & 1.9 $\times$ 10$^{-11}$ & -2.903723 & 757 & 1.5 $\times$ 10$^{-12}$ & -2.903723 \\
 & 14 & 6286 & 876 & 9.2 $\times$ 10$^{-14}$ & -2.903724\tnote{a}  & 1315 & 1.3 $\times$ 10$^{-15}$ & -2.903724\tnote{b}  \\
3 & 3 & 201 & 74 & 4.2 $\times$ 10$^{-8}$ & -2.896009 & 104 & 4.0 $\times$ 10$^{-9}$ & -2.896467 \\
 & 6 & 1338 & 292 & 3.0 $\times$ 10$^{-11}$ & -2.903488 & 459 & 1.6 $\times$ 10$^{-13}$ & -2.903501 \\
 & 10 & 5710 & 777 & 1.6 $\times$ 10$^{-13}$ & -2.903724\tnote{c}  & 1240 & 2.4 $\times$ 10$^{-15}$ & -2.903724\tnote{d}  \\
 & 14 & 15106 & 1184 & 7.7 $\times$ 10$^{-15}$ & -2.903724\tnote{e}  & 1859 & 1.0 $\times$ 10$^{-16}$ & -2.903724\tnote{f}  \\
\hline
\end{tabular} \label{tab2}
\begin{tablenotes}
\item[a] -2.903723863.
\item[b] -2.903723888.
\item[c] -2.903723618.
\item[d] -2.903723698.
\item[e] -2.903724102.
\item[f] -2.903724118.
\end{tablenotes}
\end{threeparttable}
\end{table}

At each order $n$, the electronic energies from the decontracted Gaussian expansions can be below the electronic energy of the FC method from Eqs. \eqref{slater} - \eqref{g12} without the Gaussian expansion, Eq. \eqref{gaussian_expansion}, in Table \ref{tab1}. One series of examples is $n=0, 1, 2$, and $3$ with $nG=14$ and 0.98 overlap integral selection threshold listed in Table \ref{tab1}. These results suggest the flexibilities of the Gaussian functions. 

As we can see from the data in Tables \ref{tab1} and \ref{tab2}, comparing with the reference values \cite{schwartz2006further,nakashima2007solving,kurokawa2008solving,nakashima2008accurately}, subchemical accuracy, 0.1 kcal mol$^{-1}$ \cite{koput1995ab} $\approx 1.6 \times 10^{-4}$ a.u., in the absolute energy can be reached at $n=1, nG = 10$, and 0.95 overlap integral selection threshold with 106 complement functions after the selection. However, enlarging the value of $nG$ to 14 with $n=1$ and 0.95 overlap integral selection threshold increases the electronic energy from -2.903621 to -2.903475, despite the number of the complement functions after the selection increases from 106 to 170. Similar behavior exists in $n=3$ with 0.95 overlap integral selection threshold, $nG=10$ to $nG=14$, the electronic energy increases from -2.903670 to -2.903510. These suggest the overlap integral selection method may not always align with the electronic energy optimizations, especially with the small threshold value, 0.95.

Enlarging the selection threshold of the overlap integrals from 0.95 to 0.98, 0.99, and 0.995 will improve the electronic energy and increase the number of the complement functions, except small scale computations, such as $n=0$ and $nG=3$. Notably, the smallest eigenvalues of the overlap matrices decrease along the enlarging of the selection thresholds. Though the $s_\mathrm{min}$ in Tables \ref{tab1} and \ref{tab2} are above the 
$1 \times 10^{-30}$ value in the canonical orthogonalization in Subsection \ref{subsection_im}, the present results suggest certain balance between computational cost and accuracy needs to be considered.

Comparing the results from the ECG method that all exponents are optimized \cite{rybak1989accurate}, the present approach can reach comparable accuracy. Namely, $n=3, nG=14$, and the overlap integral selection threshold 0.98,  $E \approx -2.903723829$ is similar to the value in Ref. \cite{rybak1989accurate}. The number  of the complement functions is much larger, 694, comparing with 100 in Ref. \cite{rybak1989accurate}. Nevertheless, we may the anticipate selection methods \cite{suzuki1998stochastic,muolo2018generalized,nakashima2020free} will reduce the number of the complement functions without optimizing all Gaussian exponents individually.

\section{Summary and outlook} \label{summary_section}
The present work proposes using the Gaussian expansion Eq. \eqref{gaussian_expansion} \cite{o1966gaussian,hehre1969self,hehre1970self, pietro1980molecular,pietro1981molecular,pietro1983molecular,lopez1987large,fernandez1988accurate,tew2005new, werner2007general} for the Slater functions from the initial wavefunction and the $1- e^{-r}$ type $g$ functions \cite{nakatsuji2022accurate,nakatsuji2024accurate,nakatsuji2024exact} to form Gaussian complement functions. Combined with the selection method based on the overlap integrals \cite{von2008trapped,von2009correlated,rakshit2012hyperspherical,mitroy2013theory,Kalaee2014,Mosegaard2018,moriya2023novel,coomar2022quantum}, we used the helium ground state as the example to demonstrate numerically, the FC method can encode the ECG wavefunctions. Subchemical accuracy \cite{koput1995ab} in absolute energy can be reached without optimizing the exponents of the Gaussian complement functions individually, except $nG=1$ the single Gaussian expansion.

For the similar accuracy, the number of the Gaussian complement functions is larger than the optimized ECG functions \cite{rybak1989accurate} or the FC methods without the Gaussian expanded complements \cite{nakatsuji2022accurate,nakatsuji2024accurate,nakatsuji2024exact}. Nevertheless, further advances of selection methods  \cite{suzuki1998stochastic,muolo2018generalized,nakashima2020free} would be of interest. Optimizing the exponents of the Slater functions \cite{nakashima2007solving,kurokawa2008solving,nakatsuji2024accurate}, as the sources for the Gaussian complements, would be another possibility to consider. Since the Gaussian expansion Eq. \eqref{gaussian_expansion} preserves the indices of the one- and two-electron distances in the Slater and Gaussian functions, applying the efficient antisymmetrization method \cite{nakashima2013efficient} to many-electron systems with the Gaussian complement functions could be worthwhile.

\bibliographystyle{pccp}
\bibliography{ref}

\end{document}